\documentclass[showpacs,aps,twocolumn]{revtex4}
\usepackage{epsfig}
\usepackage{graphicx}
\usepackage{amsmath,amssymb,amsfonts}
\usepackage{array}
\usepackage{url}
\usepackage{multirow}
\usepackage{float}
\usepackage{lineno}
\usepackage{xspace}
\usepackage{ulem}

\newcommand{\bef}{\begin{figure}}
\newcommand{\eef}{\end{figure}}
\newcommand{\bc}{\begin{center}}
\newcommand{\ec}{\end{center}}

\newcommand{\be}{\begin{equation}}
\newcommand{\ee}{\end{equation}}
\newcommand{\bea}{\begin{eqnarray}}
\newcommand{\eea}{\end{eqnarray}}

\def\ba{\begin{eqnarray}}
\def\ea{\end{eqnarray}}

\usepackage[usenames,dvipsnames]{color}
\definecolor{darkblue}{RGB}{0,0,196}
\usepackage[colorlinks=true,linkcolor=darkblue,citecolor=darkblue,urlcolor=darkblue]{hyperref}

\begin{document}
\title{Understanding the QCD medium by the diffusion of charm quarks using a Color String Percolation Model}
 
\author{Kangkan Goswami}
\author{Dushmanta Sahu}
\author{Raghunath Sahoo\footnote{Corresponding Author Email: Raghunath.Sahoo@cern.ch}}
\affiliation{Department of Physics, Indian Institute of Technology Indore, Simrol, Indore 453552, India}

\begin{abstract}
We study the drag and diffusion coefficients of the charm quark in the deconfined matter produced in the ultra-relativistic collisions by taking the Color String Percolation Model (CSPM) approach. CSPM, being a QCD-inspired model, can give us essential information about the hot and dense system produced in ultra-relativistic collisions. With the information on initial percolation temperature and percolation density, we estimate the relaxation time ($\tau_{c}$), drag coefficient ($\gamma$), transverse momentum diffusion coefficient ($B_{0}$), and spatial diffusion coefficient ($D_{s}$) of charm quark inside a deconfined medium. Finally, we compare the obtained results with lattice QCD and with various other theoretical models. A good agreement can be observed between the results obtained from CSPM and lattice QCD.
\end{abstract}
\date{\today}
\maketitle

\section{Introduction}
\label{intro}
Heavy quarks, such as the charm and bottom quarks, can be used as essential probes to study the initial phase of the system's evolution in an ultra-relativistic collision \cite{Xu:2017hgt, Prino:2016cni, Aarts:2016hap, Greco:2017rro}. As the charm and bottom quarks have very high masses compared to the lighter quarks, they are produced relatively early in the evolution of the system. The masses of the heavy quarks are significantly larger than that of the temperature of quark-gluon plasma (QGP), which means that the probability of the charm and bottom quarks getting created or annihilated during the deconfined phase is much less than that of the lighter quarks and gluons. As a result, the heavy quarks witness the whole space-time evolution of the system. These heavy quarks interact with the hot and dense medium formed in the ultra-relativistic collisions, because of which their momentum spectra get modified. However, these interactions do not completely thermalize heavy quarks with intermediate and high transverse momentum ($p_{\rm T}$). Thus, they can carry out important information about the initial stages of the expanding fireball.

The energy loss mechanism of the heavy quarks in the medium is very distinct from that of the lighter quarks. For light quark jets, the leading mechanism for energy loss is due to the gluon radiation \cite{Baier:1996kr}. On the contrary, the gluon radiation mechanism will be suppressed for heavy quarks \cite{Dokshitzer:2001zm}, and their energy loss will be mainly due to the elastic collisions with the lighter quarks in the medium \cite{Moore:2004tg}. Moreover, the energy of a heavy quark is not changed too much from collisions with a light quark; thus, the thermalization time of the heavy quark will be substantially larger than that of the lighter quarks. It is worth mentioning that the QGP lifetime is estimated to be in the order of 4-5 fm/c \cite{Heinz:2002gs} at the Relativistic Heavy Ion Collider (RHIC). Similarly, at the Large Hadron Collider (LHC), it is around 10-12 fm/c \cite{Foka:2016vta}. On the other hand, thermalization time for charm quark is calculated to be in the order of 10-15 fm/c, and for bottom quarks, 25-30 fm/c \cite{Moore:2004tg, vanHees:2005wb, Cao:2011et}. The charm quark interaction with the thermalized lighter quarks and gluons in the medium will lead to a Brownian motion, which can be explained by the Fokker-Planck transport equation. Hence, information about the interaction of a heavy quark in the deconfined medium is preserved within its drag and diffusion coefficients, which can be estimated by solving the Fokker-Planck equation. While traversing through the medium, the average momentum of heavy quarks gets modified, which is incorporated in their drag coefficient. Their momentum distribution is also broadened, which is embedded in their momentum diffusion coefficient. These coefficients vary with the temperature of the expanding matter, and in principle, they can give essential information about the systems formed in high-energy collisions.

 Apart from the drag and momentum diffusion coefficients, the spatial diffusion coefficient ($D_{s}$) is also of enormous interest to the scientific community. Theoretically, the heavy quark spatial diffusion coefficient  has been calculated from perturbative QCD (pQCD) \cite{vanHees:2004gq} and also from Anti-de Sitter/Conformal Field Theory (AdS/CFT) correspondence \cite{Gubser:2006qh}. AdS/CFT calculations give the value of $2\pi TD_{s}$ = 1 in the strong coupling limit \cite{Kovtun:2003wp}. As observed from several works in the literature, there is a minimum for this parameter approaching the AdS/CFT value near the critical temperature. In Ref.~\cite{Scardina:2017ipo}, the authors have estimated the charm quark spatial diffusion coefficient from $D$ meson spectra at energies available at RHIC and LHC. The heavy quark momentum and spatial diffusion coefficient have also been obtained from lattice QCD by investigating a color electric field correlation function using Monte-Carlo techniques \cite{Banerjee:2011ra}. Recently, Brambilla et al. have reported some progress towards the computation of heavy quark momentum diffusion coefficient from two chromoelectric fields correlator attached to a Polyakov loop in SU(3) gauge theory \cite{Brambilla:2020siz}. By studying the diffusion coefficient as a function of temperature, they show that the results from lattice QCD agree quite well with that of the next-to-leading order (NLO) perturbative results. The effect of viscosity on the diffusion coefficient has also been studied by taking a semi-QGP matrix model \cite{Singh:2019cwi}, where the spatial diffusion coefficient is observed to be decreasing with the increase in shear and bulk viscosities. In addition, the diffusion of $D$ meson in the hot and dense hadronic matter has been studied for both zero and finite baryochemical potential cases \cite{Ozvenchuk:2014rpa}. A strong baryochemical potential dependency has been observed in the estimation of relaxation time for $D$ meson; for higher $\mu_{B}$, the relaxation time is observed to be reduced by a factor of 2-3.

 It has been observed that the Color String Percolation Model (CSPM) can qualitatively explain the observed increase in the relative $J/\psi$ yield with respect to relative charged-particle multiplicity in pp collisions at $\sqrt{s}$ = 7 TeV \cite{Ferreiro:2012fb}. This motivates us to look into heavy flavor dynamics in the deconfined medium by using CSPM. In this work, we study the relaxation time, drag, and momentum diffusion coefficient of the charm quark, along with $D_{s}$ as functions of initial percolation temperature by taking the CSPM approach. The paper is organized as follows. Sec.~\ref{formulation} encompasses the CSPM framework along with the formalism of the drag and diffusion coefficients of the charm quark. In Sec.~\ref{res}, the results obtained using the formulation are discussed. Finally, Sec.~\ref{con} concludes the findings of this study.

\section{Formulation}
\label{formulation}
\subsection{Color String Percolation Model}

\begin{figure}[ht!]
\begin{center}
\includegraphics[scale = 0.25]{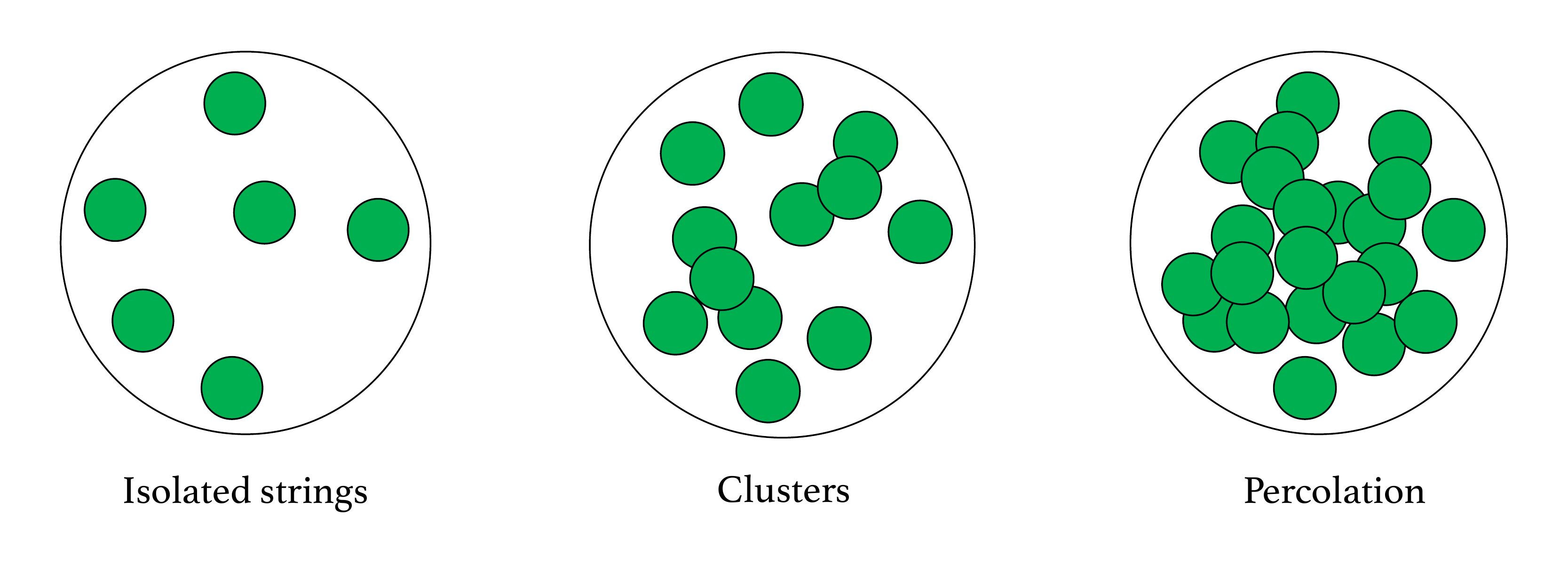}
\caption{(Color Online) Percolation of colored strings in the transverse space.}
\label{fig1}
\end{center}
\end{figure}

The color string percolation model is a QCD-inspired, well-established model to explain the multi-particle production in an ultra-relativistic collision \cite{Braun:2015eoa}. The model has been used to estimate various thermodynamic and transport properties of the matter formed in ultra-relativistic collisions, and the results have been found to be consistent with other existing models such as lattice QCD \cite{Sahu:2020mzo,Mishra:2021yer,Sahu:2022bzg}. In this model, the production of particles can be understood as a consequence of color strings stretching between the partons of the target and the projectile. In transverse space, these strings occupy some finite area. With an increase in collision energy and the number of colliding partons, the number of strings grows, and they start to overlap, forming colored clusters in the transverse space. A macroscopic cluster appears after a certain critical string density ($\xi_{\rm c}$), where 50$\%$ of the transverse space is occupied by the clusters. This marks the percolation phase transition \cite{Braun:2015eoa}, which is shown schematically in Fig. \ref{fig1}. According to Scwhinger's string breaking mechanism, these strings stretch and break to produce color-neutral quark and antiquark pairs. These particles finally hadronize to produce the final state hadrons \cite{Braun:2015eoa}. 

In a 2D percolation theory, the total multiplicity due to $N_{\rm s}$ number of overlapping strings is given as \cite{Braun:1999hv},
\begin{equation}
\label{eq1}
    \mu_{N_{s}}/\mu_1 = \sum_{n=1}^{N_{\rm s}}\sqrt{n}(S_{\rm n}/S_{\rm 1})
\end{equation}
where, $n$ is any interger from 1 to $N_{\rm s}$, $S_{1}$ is the transverse area of a single string given by $S_{\rm 1} = \pi r_{\rm 0}^{2}$ with $r_{\rm 0}$ being the single string radius $\simeq$ 0.2 fm, and $S_{\rm n}$ is the transverse area occupied by the overlapping strings.

Similarly, for mean transverse momentum squared, $\langle p^2_{\rm T} \rangle$, we can write,
\begin{equation}
\label{eq2}
    \langle p^{2}_{\rm T} \rangle/\langle p^{2}_{\rm T} \rangle_{1} = \frac{N_{\rm s}}{\sum_{n=1}^{N_{\rm s}}\sqrt{n}(S_{\rm n}/S_{\rm 1})}.
\end{equation}

When the strings in the transverse plane are just touching each other such that, $S_{\rm n} = nS_{1}$, the total multiplicity and the mean transverse momentum squared changes as $\mu_{\rm n} = n\mu_{\rm 1}$ and $\langle p^{2}_{\rm T} \rangle_{n} = \langle p^{2}_{\rm T} \rangle_{1}$ respectively. Another possible scenario is that all the strings overlap perfectly over each other such that, $S_{\rm n} = S_{1}$. In this case, we get $\mu_{\rm n} = \sqrt{n}\mu_{\rm 1}$ and $\langle p^{2}_{\rm T} \rangle_{n} = \sqrt{n}\langle p^{2}_{\rm T} \rangle_{1}$. This gives us the case where the multiplicity is maximally suppressed and the mean transverse momentum squared is maximally enhanced. From this, we can write a relation between multiplicity and mean transverse momentum squared, $\mu_{\rm n}\langle p^{2}_{\rm T} \rangle_{\rm n} = n\mu_{\rm 1}\langle p^{2}_{\rm T} \rangle_{1}$, which implies conservation of total transverse momentum.

Now, assuming the transverse nuclear overlap area to be $S$ and density $\rho$, we introduce a dimensionless percolation density parameter, $\xi$,  is given by~\cite{Braun:2015eoa}, 
\begin{equation}
\label{eq3}
\xi = \rho S_{\rm 1} = \frac{N_{\rm s}S_{1}}{S},
\end{equation}

In the thermodynamic limit, the number of the strings
$N_{s} \rightarrow\infty $ while $\xi$ is  fixed, and the distribution of the overlaps of $n$ strings is Poissonian with a 
mean of value $\xi$, 
\begin{equation}
\label{eq4}
p_n= \frac{\xi^n}{n!}e^{-\xi}.
\end{equation}

From the above expression, the fraction of the total area occupied by the strings can be given by,
\begin{equation}
\label{eq5}
\sum_{n=1}p_n=1-e^{-\xi}.
\end{equation}

Dividing the above equation by $\xi$, we get the compression factor. Furthermore, according to the CSPM approach, the
multiplicity becomes damped as a result of overlapping by the square root of the compression factor, which can be written as,
\begin{equation}
\label{eq6}
    \mu_n/\mu_1 = N_{\rm s}\sqrt{\frac{1-e^{-\xi}}{\xi}}.
\end{equation}

Hence, finally the damping factor or the color suppression factor is given as \cite{Braun:2015eoa},
\begin{equation}
\label{eq7}
F(\xi)=\sqrt{\frac{1-e^{-\xi}}{\xi}}.
\end{equation}


In ultra-relativistic heavy-ion collisions, thermalization can occur through Hawking-Unruh effect \cite{Hawking:1975vcx,Unruh:1976db}, where fast thermalization happens with the existence of an event horizon caused by the rapid deceleration of the colliding nuclei \cite{Castorina:2007eb}. In CSPM, the intense color field inside the large cluster causes deceleration of the $q\bar q$ pair, which can be perceived as a thermal temperature due to the Hawking-Unruh effect. This suggests that the radiation temperature can be determined by the transverse extension of the color flux tube in terms of the string tension.

The initial temperature of the percolation cluster is expressed in terms of $F(\xi)$. The Schwinger distribution for massless particles can be expressed in terms of $p_{\rm T}^2$ as \cite{Hirsch:2018pqm, Braun:2015eoa, cywong:1994, Schwinger:1962tp}:
\begin{equation}
\label{eq8}
    dn/dp_{\rm T}^2 \sim {\rm exp} (-\pi p_{\rm T}^2/x^2)
\end{equation}
where $\langle x^2 \rangle$ is the average value of string tension. As the chromoelectric field is not constant, the tension of the macroscopic cluster fluctuates around its mean value. Due to these fluctuations, we get a Gaussian distribution of the string tension given as, 

\begin{equation}
\label{eq9}
    \frac{dn}{dp_{\rm T}^2} \sim \sqrt{\frac{2}{\langle x^2 \rangle}} \int_{0}^{\infty} dx   \exp\left(- \frac{x^2}{2\langle x^2 \rangle} \right) \exp\left( -\pi \frac{p_{\rm T}^2}{x^2} \right)
\end{equation}
which in turn gives rise to a thermal distribution,

\begin{equation}
\label{eq10}
    \frac{dn}{dp_{\rm T}^2} \sim \exp \left(  -p_{\rm T}^2 \sqrt{\frac{2 \pi}{\langle x^2 \rangle}} \right).
\end{equation}
From Eq.\ref{eq8} and \ref{eq10}, the initial temperature in terms of $F(\xi)$ can be expressed as \cite{Hirsch:2018pqm,Sahoo:2017umy,Mishra:2020epq},
\begin{equation}
\label{eq11}
T(\xi) = \sqrt{\frac{\langle p^{2}_{\rm  T}\rangle_{1}}{2F(\xi)}},
\end{equation}
where, $\langle x^{2} \rangle = \pi \langle p_{\rm T}^{2}\rangle_{1}/F(\xi)$. By using $T_{c} = 167.7\pm2.8$ MeV \cite{Braun-Munzinger:2003htr,Becattini:2010sk} and $\xi_{\rm c} \sim 1.2$ \cite{Braun:2015eoa}, we can get the single string squared average transverse momentum, $ \sqrt{ \langle p^{2}_{\rm T}\rangle_{1}} = 207.2\pm3.3$~MeV. Using this value in Eq. \ref{eq7}, we can get the initial temperature for different $F({\xi})$ values.

With this information, let's estimate the drag and diffusion coefficients within CSPM formalism.

\subsection{Drag and diffusion coefficients}

To study the interaction of charm quark with the thermal quarks inside the deconfined medium, we take advantage of the Fokker-Planck transport equation which can be written as,
\begin{equation}
\label{eq12}
\frac{\partial f(t,p)}{\partial t} = \frac{\partial}{\partial p^{i}}\bigg[ (A^{i} f(t,p)) + \frac{\partial}{\partial p^{j}} ( B^{ij} f(t,p))\bigg].
\end{equation}
Here, $f(t,p)$ is the time evolution phase-space distribution of charm quarks. The kernels $A^{i}$ and $B^{ij}$  are given by \cite{Ozvenchuk:2014rpa},
\begin{equation}
\label{eq13}
A^{i} = \int_{}^{}dk~ \omega(p,k)k^{i},
\end{equation}

\begin{equation}
\label{eq14}
B^{ij} = \frac{1}{2} \int_{}^{}dk~ \omega(p,k)k^{i}k^{j},
\end{equation}
where $\omega(p,k)$ is the collision rate of the charm quark, with initial momenta $p$ and final momenta $(p-k)$, with the transferred momenta $k$. $i,j$ = 1,2,3 are the spatial indices. In the low transverse momentum limit $(p\to 0)$, the kernels reduce to \cite{Ozvenchuk:2014rpa},
\begin{equation}
\label{eq15}
A_{i} = \gamma p_{i},
\end{equation}

\begin{equation}
\label{eq16}
B_{ij} = B_{0}P_{ij}^{\bot}+B_{1}P_{ij}^{\parallel},
\end{equation}
where $\gamma$ is the drag coefficient, $B_{0}$ is the transverse momentum diffusion coefficient, and $B_{1}$ is the longitudinal momentum diffusion coefficient. $P_{ij}^{\parallel}$ and $P_{ij}^{\bot}$ are the longitudinal and transverse projection operators respectively. This drag and diffusion coefficient definition is valid only when the transferred energy is small. This is possible in the case of charm quark being non-relativistic. The NRQCD model assumes the heavy particles to be non-relativistic, and it describes the inclusive production and decay of quarkonia along with the $S$-wave charmonium production at high transverse momentum \cite{Bodwin:1994jh}. Going in line with this, we can assume the charm quark inside the deconfined medium to be non-relativistic. Thus the assumptions of Eq.\ref{eq15} and Eq.\ref{eq16} hold true in this case.

The average momentum of a particle at a given time is given by \cite{Torres-Rincon:2013nfa},
\begin{equation}
\label{eq17}
\langle p \rangle = \frac{ \int_{-\infty}^{\infty} dp ~ p ~f(t,p)}{\int_{-\infty}^{\infty} dp~f(t,p)} = p_{0}~e^{-\frac{t}{\tau}}
\end{equation}
where, $\tau$ is the relaxation time and $p_{\rm 0}$ is the initial momentum of the particle. Relaxation time of a quark in the hot QCD medium is defined as the time accounting for the exponential decay of the average momentum. In other words, the average time interval between two successive collisions can be termed as the relaxation time of the system. Assuming the velocity of the thermal quarks to be nearly the speed of light, in natural units we can take the relaxation time to be equal to the mean free path of the system.  The mean free path ($\lambda$) is inversely proportional to the number density ($n$) of the system and the scattering cross-section ($\sigma$) between the particles \cite{Sahoo:2017umy,Braun:2015eoa},
\begin{equation}
\label{eq18}
\lambda = \frac{1}{n\sigma}.
\end{equation}
In CSPM formalism, the number density is expressed as the effective number of color sources per unit volume:
\begin{equation}
\label{eq19}
n = \frac{N_{\rm sources}}{S L},
\end{equation}

where $S$ is the nuclear overlap area and $L$ is the longitudinal extension of a string $\sim$ 1 fm. This particular string length is chosen because no new quark-antiquark pairs can form if the separation between the gluons is less than 1 fm. In addition, the colliding nuclei are Lorentz contracted, making their longitudinal dimensions almost negligible. Thus, we take the lower limit of the string length, which is standard in CSPM studies.

The number of color sources can be defined as the transverse area occupied by the strings divided by the area of an effective string:
\begin{equation}
\label{eq20}
N_{\rm sources} = \frac{(1-e^{-\xi})S_{\rm N}}{S_{\rm 1}F(\xi)}.
\end{equation}
Thus, the number density becomes,
\begin{equation}
\label{eq21}
n = \frac{(1-e^{-\xi})}{S_{\rm 1}F(\xi)L}.
\end{equation}
The cross-section ($\sigma$) can also be expressed as the transverse area of the effective strings $  S_{\rm 1}F(\xi)$. Finally, using Eq.\ref{eq18} and \ref{eq21} we can write,
\begin{equation}
\label{eq22}
\lambda = \frac{L}{(1-e^{-\xi})}.
\end{equation}
The above equation can be used to define the relaxation time of the deconfined medium, $\tau = \lambda$ \cite{Rapp:2008qc}.

Now, for heavier quarks like charm quark, the particle dependent relaxation time can be expressed by \cite{Petreczky:2005nh},
\begin{equation}
\label{eq23}
\tau_{\rm c} = \frac{m_{\rm c}}{T}\tau
\end{equation}
\begin{equation}
\label{eq24}
\Rightarrow \tau_{\rm c} = \frac{m_{\rm c}}{T}\frac{L}{(1-e^{-\xi})},
\end{equation}
where $m_{\rm c}$ is the mass of the charm quark, $m_{c}$ $\simeq$ 1.275 GeV. Due to the dependency of relaxation time on mass of the particle, at any temperature, we expect the relaxation time of the charm quark to be significantly higher than that of the lighter quarks. This, in turn, affects the elliptic flow of charm quark to be smaller than that of the light hadrons \cite{Scardina:2017ipo, vanHees:2004gq}. 

The drag coefficient or drag force ($\gamma$), which incorporates average momentum change, is inversely related to the thermal relaxation time of the particle, $\gamma = \frac{1}{\tau_{\rm c}}$. From the Einstein's relation, the transverse momentum diffusion coefficient for charm quark, which accounts for broadening of the final momentum distribution, is related to the drag coefficient as \cite{Tolos:2016slr},
\begin{equation}
\label{eq25}
B_{\rm 0} = \gamma Tm_{\rm c}
\end{equation}
\begin{equation}
\label{eq26}
\Rightarrow B_{\rm 0} = \frac{(1-e^{-\xi})T^{2}}{L}.
\end{equation}

The diffusion coefficient $(D_{s})$ in position space can also be introduced. It can be estimated by starting a particle at position and time $x = 0$ and $t = 0$, and finding the mean squared position at a later time \cite{Torres-Rincon:2013nfa},
\begin{equation}
\label{eq27}
\langle (x(t) - x(0))^{2} \rangle = 2D_{s}t.
\end{equation}
Here, we can see that $D_{s}$ is the measure of speed of diffusion in space and is called as``Einstein relation". In static limit, the spatial diffusion coefficient is given by the expression \cite{Torres-Rincon:2013nfa,DsBook},
\begin{equation}
\label{eq28}
D_{\rm s} = \frac{T}{m_{\rm c }\gamma}
\end{equation}
\begin{equation}
\label{eq29}
\Rightarrow D_{\rm s} = \frac{L}{(1-e^{-\xi})}.
\end{equation}

By taking inputs from the CSPM formalism and using the above formulation, we have estimated the relaxation time, drag, and diffusion coefficients of the charm quarks inside the deconfined medium, which are discussed in the following section.

\section{Results and Discussion}
\label{res}

\begin{figure}[ht!]
\begin{center}
\includegraphics[scale = 0.40]{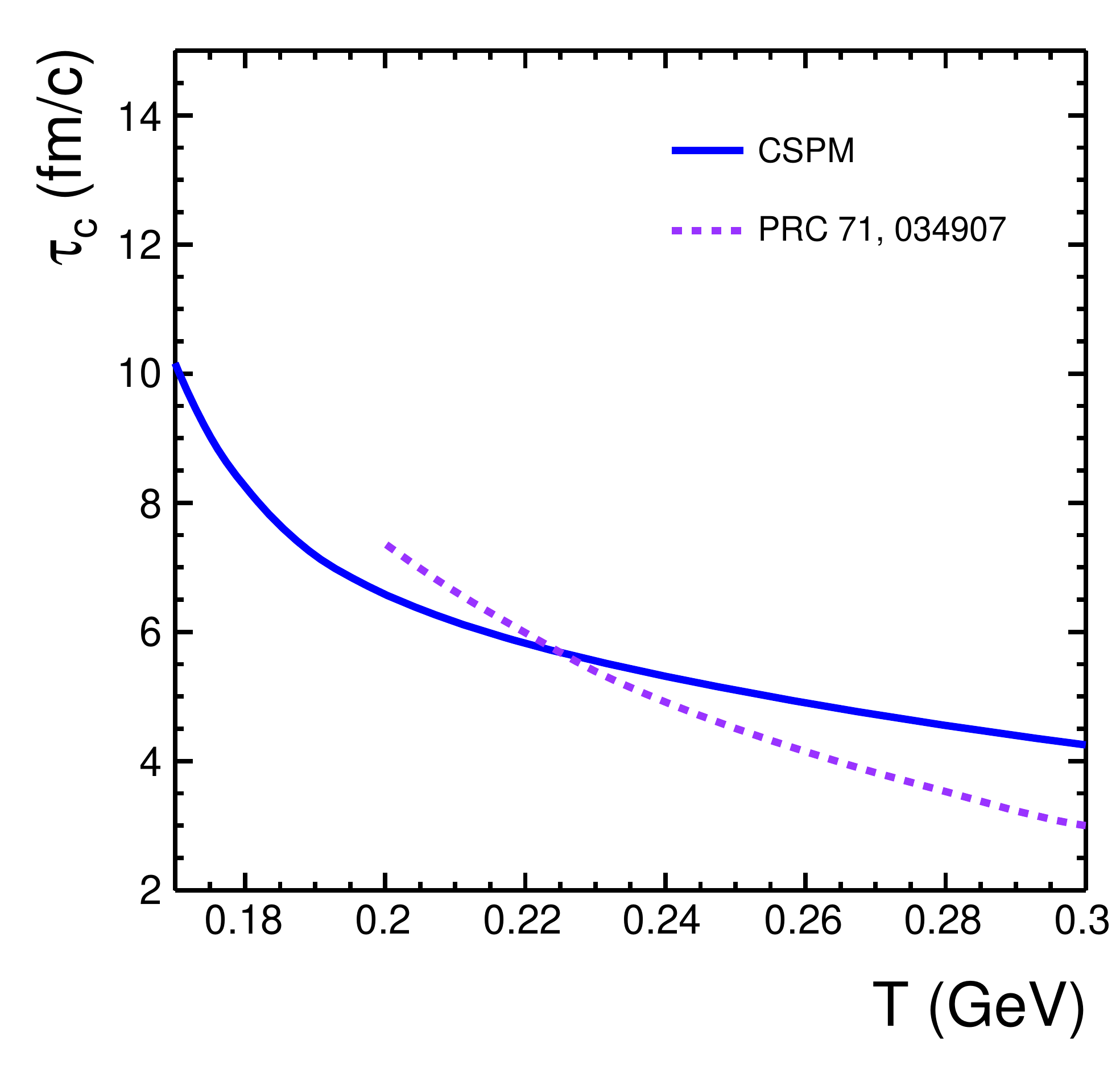}
\caption{(Color Online) Relaxation time of charm quarks as a function of initial percolation temperature. The dotted violet line is from pQCD estimations with resonant particles \cite{vanHees:2004gq}.}
\label{fig2}
\end{center}
\end{figure}

To better understand the systems produced in the ultra-relativistic collisions at RHIC and LHC, we have studied the relaxation time, drag, and diffusion coefficients of such matter using the CSPM. The results are contrasted with the results obtained from various other models. Firstly, the charm quark relaxation time is plotted as a function of the initial temperature in fig.~\ref{fig2}. $\tau_{\rm c}$ signifies the amount of time it takes for the charm particle to lose its total momentum by interaction with other quarks and gluons while traversing through the QGP medium. We observe that the relaxation time decreases with the increase in initial temperature. At temperature nearing the critical value, the relaxation time of the charm quark is around 10 fm/c. With the rise in initial temperature, the charm quark relaxation time decreases, and at 300 MeV, it becomes about 4 fm/c. This is comparable with the result obtained in ref. \cite{vanHees:2004gq}, where a pQCD approach has been taken considering resonant particles' inclusion inside the medium. The charm quark relaxation time is lower for a denser system such as the one formed in most central heavy-ion collisions. In comparison, $\tau_{\rm c}$ is relatively higher for a less dense medium, such as the one formed in high multiplicity pp collisions or peripheral heavy-ion collisions. This is because the charm quark will thermalize faster in a denser medium than in a less dense one due to substantially more interactions in the medium.

\begin{figure}[ht!]
\begin{center}
\includegraphics[scale = 0.40]{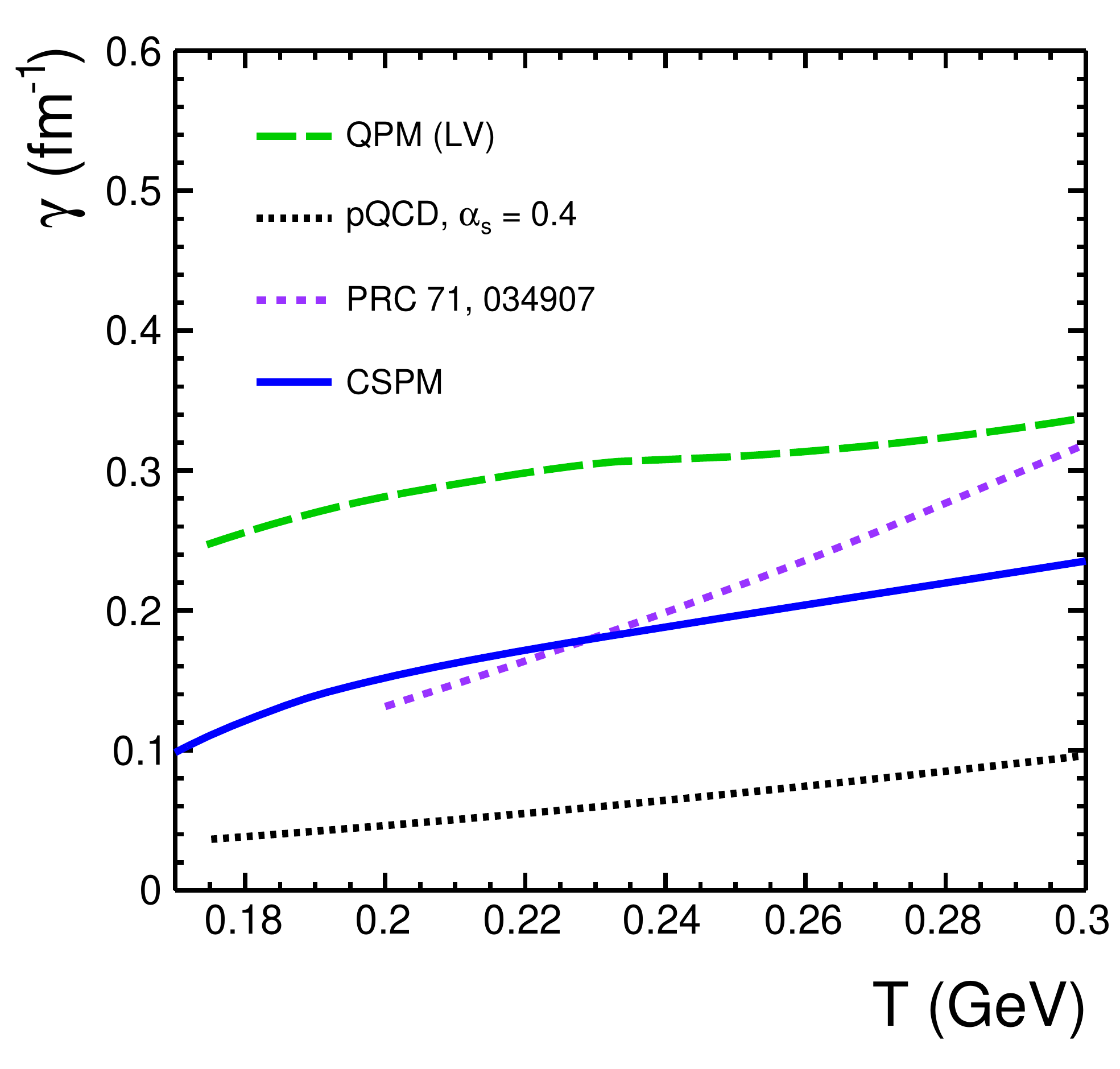}
\caption{(Color Online) Drag coefficient of charm quark as a function of temperature. The dashed green line is from the quasi-particle model \cite{Scardina:2017ipo}, the dotted black and violet lines are from pQCD estimations without and with the inclusion of resonant particles, respectively \cite{vanHees:2004gq}.}
\label{fig3}
\end{center}
\end{figure}

By taking input from the charm quark relaxation time, we estimate the drag force of the charm quark inside the medium, which is the inverse of relaxation time. Fig.~\ref{fig3} shows the drag force or drag coefficient of charm quark in the deconfined medium as a function of temperature. We observe an increasing trend with the increase in temperature. This suggests that the drag on the charm quark in denser systems, such as those produced in most central heavy-ion collisions, will be more than that in the peripheral collisions. For comparison, we have also plotted the drag force of charm quark estimated from the quasi-particle model \cite{Scardina:2017ipo} and pQCD \cite{vanHees:2004gq}. The drag coefficient calculated from the quasi-particle model is a little higher than the CSPM estimation, whereas $\gamma$ obtained from pQCD lies below the CSPM results. But there is a considerable agreement between the CSPM and the pQCD results with the inclusion of resonant particles inside the medium \cite{vanHees:2004gq}.

\begin{figure}[ht!]
\begin{center}
\includegraphics[scale = 0.40]{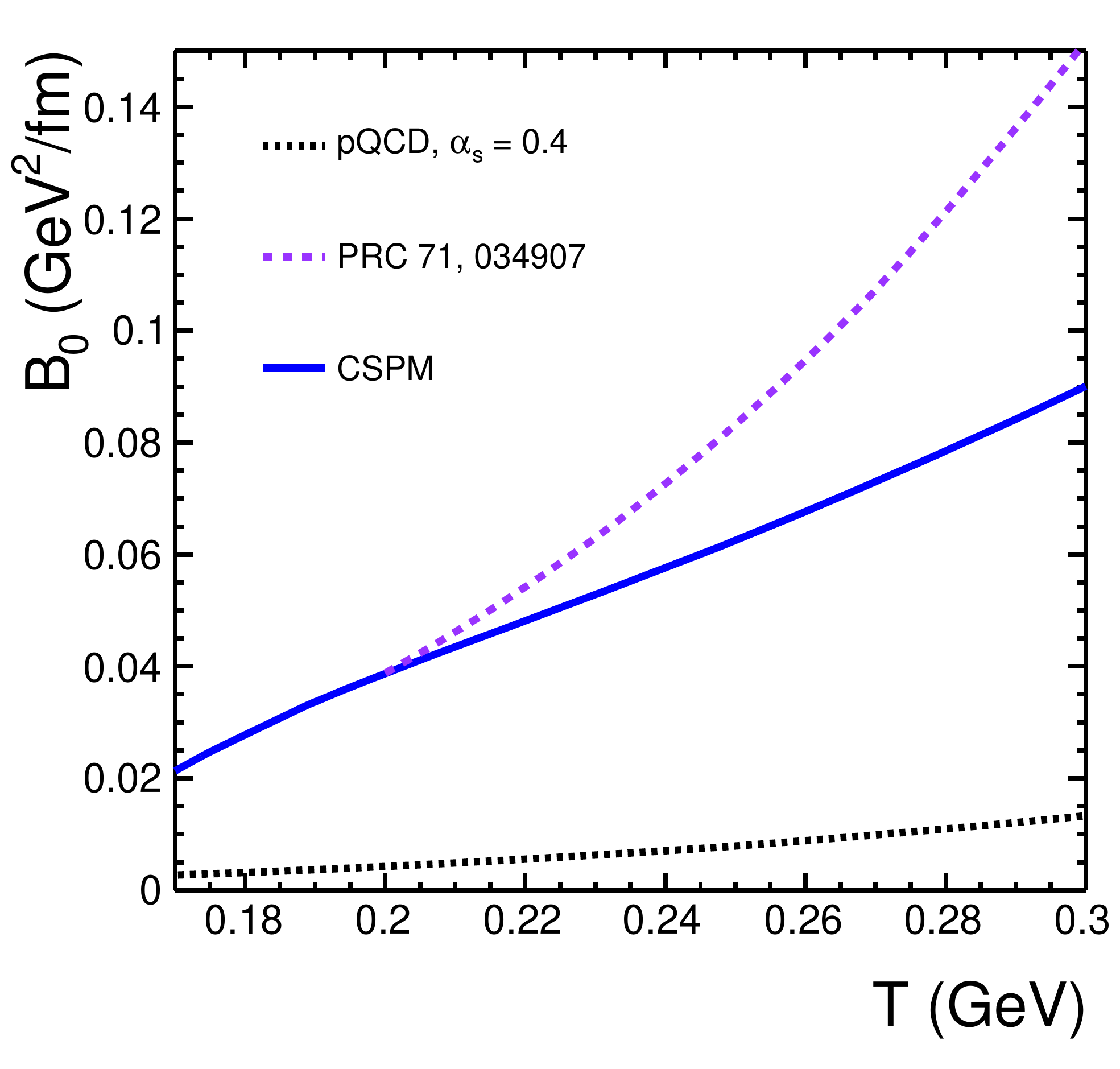}
\caption{(Color Online) Transverse momentum diffusion coefficient of charm quark as a function of temperature. The dotted black and violet lines are from pQCD estimations without and with the inclusion of resonant particles, respectively \cite{vanHees:2004gq}.}
\label{fig4}
\end{center}
\end{figure}

From Einstein's relation, the drag and momentum diffusion coefficients are related by the Eq.~\ref{eq25}. Fig.~\ref{fig4} shows the transverse momentum diffusion coefficient as a function of temperature. We observe that $B_{0}$ increases linearly with an increase in temperature, suggesting that for high temperature, the momentum broadening of the charm quarks will be higher. We have also plotted the transverse momentum diffusion coefficient from pQCD to compare our results. The charm quark momentum diffusion from CSPM estimation is substantially larger than that of pQCD calculation. This momentum diffusion of charm quarks, in principle, can affect the elliptic flow of the final state charmed hadrons. It is worthy to note that, in the hadronic medium, $D_{\rm 0}$ meson will diffuse considerably larger than $J/\psi$ \cite{Mitra:2014ipa}. This essentially results in a suppressed $v_{\rm 2}$ of $D_{\rm 0}$. But, because $J/\psi$ remains largely undiffused in the hadronic phase, the elliptic flow of $J/\psi$ will give unfiltered information about the deconfined phase, making $J/\psi$ a cleaner probe to study QGP.

\begin{figure}[ht!]
\centering
\includegraphics[scale = 0.40]{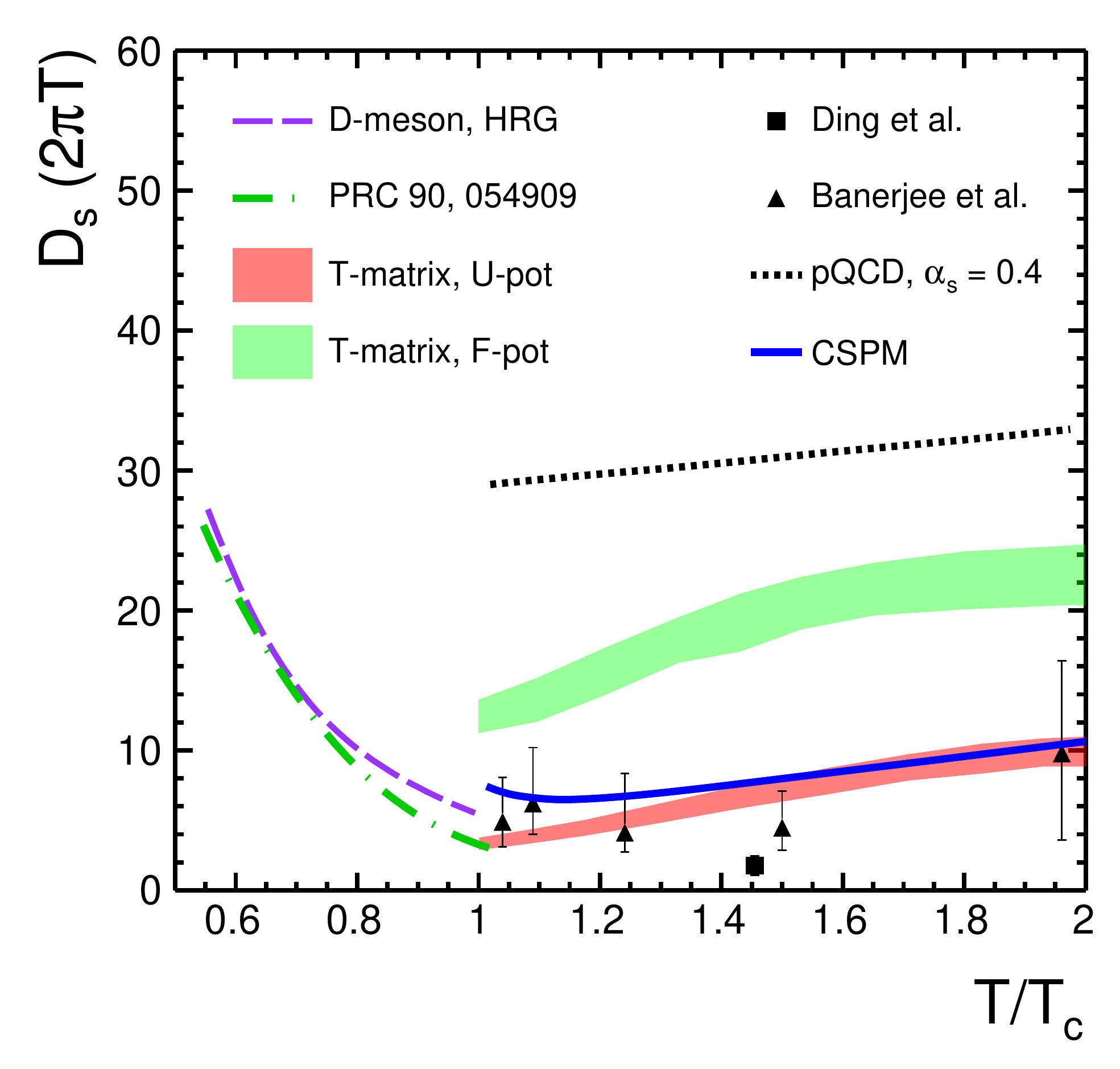}
\caption{(Color Online) The spatial diffusion coefficient as a function of temperature scaled with critical temperature. The red and green bands are the results obtained from T-matrix calculations for U-pot and F-pot respectively \cite{Riek:2010fk}. The black dotted line is from pQCD calculations \cite{vanHees:2004gq}, the triangle \cite{Banerjee:2011ra} and rectangle \cite{Ding:2012sp} markers are from lattice QCD estimations. The violet dashed line \cite{Tolos:2013kva} and the dashed-dotted green line \cite{Ozvenchuk:2014rpa} are the results from D meson diffusion in a hadron gas.}
\label{fig5}
\end{figure}

Finally, in fig.~\ref{fig5} we have plotted the spatial diffusion coefficient as a function of temperature scaled with critical temperature. Expressing the transport coefficients in the units of thermal wavelength is very convenient; thus, we have multiplied a factor of $(2\pi T)$, making the diffusion coefficient dimensionless. This dimensionless quantity can characterize the coupling strength of the charm quark to the thermal medium. We have compared our results with various other results that already exist in the literature, such as estimations from T-matrix \cite{Riek:2010fk}, pQCD \cite{vanHees:2004gq} and lattice QCD \cite{Banerjee:2011ra, Ding:2012sp} calculations. Interesting results have also been observed from the D meson diffusion in the hadron gas \cite{Ozvenchuk:2014rpa,Tolos:2013kva}. As the temperature approaches $T_{\rm c}$, the D meson diffusion decreases rapidly and is almost comparable with that of charm quark diffusion values at the critical temperature, showing almost a smooth transition from hadronic to partonic phase. From AdS/CFT calculations, the shear viscosity to entropy density ratio gives the minimum value of $1/4\pi$ \cite{Kovtun:2004de}. When we study the change of $\eta/s$ with temperature \cite{Sahu:2020mzo}, we observe that $\eta/s$ decreases with an increase in temperature and becomes minimum at $T=T_{c}$, and then again starts increasing with temperature. In addition, from the study of $\zeta/s$ as a function of temperature \cite{Sahu:2020mzo}, we observe that at $T=T_{c}$ the bulk viscosity to entropy density becomes minimum and almost close to zero. Similarly, AdS/CFT calculation gives a minimum of $D_{s}(2\pi T)$ $\sim$ 1 \cite{Kovtun:2003wp} at the critical temperature. From fig.~\ref{fig5} we can see that the value of $D_{s}(2\pi T)$ approaches this minima near $T_{\rm c}$. Furthermore, we observe a good agreement between the CSPM and other model estimations. 

This behavior of spatial diffusion coefficient can be understood in terms of the interaction strength in the system. In the hadronic phase, at a lower temperature, the interaction will be less; thus, the spatial diffusion coefficient is higher. As the temperature increases and we go towards the critical temperature, the interaction also increases, resulting in lower values of  $D_{s}(2\pi T)$. At $T=T_{\rm c}$, the interaction will be maximum due to the onset of the QGP medium, which corresponds to the minimum of $D_{s}(2\pi T)$. After this, the spatial diffusion coefficient increases again with an increase in temperature. This is because, at a higher temperature, the patrons will be asymptotically free, resulting in a lower interaction strength.

\section{Conclusion}
\label{con}
In this work, for the first time, we study the relaxation time, drag, and diffusion coefficients of charm quark in the deconfined medium by taking the color string percolation approach. Our result is comparable with the pQCD approach, where the resonant heavy-light quark interactions are introduced. The observations from the spatial diffusion coefficient state that there is a minimum at the phase transition. We observe that our findings from the CSPM agree with the results obtained from lattice QCD for the spatial diffusion coefficient. 

In the view of ALICE RUN-3 going towards high luminosity, the heavy flavor sector will be of particular interest. With higher statistics, the study of hadron production containing charm and bottom quarks can be done with 
high precision for a broad multiplicity range. Our work, along with other theoretical works will hopefully help to understand heavy flavor dynamics in ultra-relativistic collisions as case studies.

\section*{Acknowledgment}
KG acknowledges the doctoral fellowship from UGC, Government of India.
The authors gratefully acknowledge the DAE-DST, Govt. of India funding under the mega-science project – “Indian participation in the ALICE experiment at CERN” bearing Project No. SR/MF/PS-02/2021-IITI (E-37123).

 \end{document}